\documentclass[twocolumn]{aastex62}
\pdfoutput=1

\newcommand\fidanka{\texttt{Fidanka} }

\usepackage{amsmath}
\usepackage{cancel}
\usepackage{float}

\shorttitle{Self Consistently Modeling NGC 2808}
\shortauthors{Boudreaux et al.}
\graphicspath{{./}{figures/}{src/figures}}

\begin{document}

\title{Chemically Self-Consistent Modeling of the Globular Cluster NGC 2808 and its Effects on the Inferred Helium Abundance of Multiple Stellar Populations}
\received{February 7, 2024}
\revised{October 7, 2024}
\accepted{November 22, 2024}

\correspondingauthor{Emily M. Boudreaux}
\email{emily.m.boudreaux.gr@dartmouth.edu, emily@boudreauxmail.com}

\author[0000-0002-2600-7513]{Emily M. Boudreaux}
\affiliation{Department of Physics and Astronomy, Dartmouth College, Hanover, NH 03755, USA}

\author[0000-0003-3096-4161]{Brian C. Chaboyer}
\affiliation{Department of Physics and Astronomy, Dartmouth College, Hanover, NH 03755, USA}

\author{Amanda Ash}
\affiliation{Department of Physics and Astronomy, University of North Georgia, Dahlonega, GA 30533, USA}

\author{Renata Edaes Hoh}
\affiliation{Department of Physics and Astronomy, Dartmouth College, Hanover, NH 03755, USA}

\author[0000-0002-2012-7215]{Gregory Feiden}
\affiliation{Department of Physics and Astronomy, University of North Georgia, Dahlonega, GA 30533, USA}

\begin{abstract}
  The helium abundances in the multiple populations that are now known to
  comprise all closely studied Milky Way globular clusters are often inferred
  by fitting isochohrones generated from stellar evolutionary models to
  globular cluster photometry. It is therefore important to build stellar
  models that are chemically self-consistent in terms of their structure,
  atmosphere, and opacity. In this work we present the first chemically
  self-consistent stellar models of the Milky Way globular cluster NGC 2808
  using MARCS model atmospheres, OPLIB high-temperature radiative opacities,
  and AESOPUS low-temperature radiative opacities. These stellar models were
  fit to the NGC 2808 photometry using \fidanka, a new software tool that was
  developed to optimally fit cluster photometry to isochrones and for population
  synthesis.  \fidanka can determine, in a relatively unbiased way, the ideal
  number of distinct populations that exist within a dataset and then fit
  isochrones to each population. We achieve this outcome through a combination of
  Bayesian Gaussian Mixture Modeling and a novel number density estimation
  algorithm. Using \fidanka and F275W-F814W photometry from the Hubble UV
  Globular Cluster Survey we find that the helium abundance of the second
  generation of stars in NGC 2808 is higher than the first generation by
  $15\pm3\%$. This is in agreement with previous studies of NGC 2808. This
  work, along with previous work by \citet{Dotter2015} focused on NGC 6752,
  demonstrates that chemically self-consistent models of globular clusters do
  not significantly alter inferred helium abundances, and are therefore unlikely
  to be worth the significant additional time investment.
\end{abstract}


\section{Introduction}\label{sec:Intro}
Globular clusters (GCs) are among the oldest observable objects in the universe
\citep{Pen11}. They are characterized by high densities with typical half-light
radii of $\le$10 pc \citep{Vanderburg2010}, and typical masses ranging from
$10^{4}$--$10^{5}$ M$_{\odot}$ \citep{Bro06} --- though some GCs are
significantly larger than these typical values \citep[e.g. $\omega$ Cen,
][]{Richer1991}. GCs provide a unique way to probe stellar evolution
\citep{Kalirai2010}, galaxy formation models \citep{Boy18,Kra05}, and dark matter
halo structure \citep{Hud18}.

The traditional view of globular clusters is that they consist of a single
stellar population (SSP, in some publications this is referred to as a Simple
Stellar Population). This view was supported by spectroscopically uniform heavy
element abundances \citep{Carretta2010, Bastian2018} across most clusters (M54
and $\omega$Cen are notable exceptions, see \citet{Marino2015} for further
details), and the lack of evidence for multiple stellar populations (MPs) in
past color-magnitude diagrams of GCs \citep[i.e.][]{Sandage1953, Alcaino1975}.
However, over the last 40 years non-trivial star-to-star light-element
abundance variations have been observed \citep[i.e.][]{Smith1987} and, in the
last two decades, it has been definitively shown that most, if not all, Milky Way
GCs have MPs \citep{Gratton2004, Gratton2012, Piotto2015}. The lack of
photometric evidence for MPs prior to the 2000, can be attributed to the more
narrow color bands available, until very recently, to ground-based photometric
surveys \citep{Milone2017}.

The prevalence of multiple populations in GCs is so distinct that the proposed
definitions for what constitutes a globular cluster now often center on the
existence of MPs \citep[e.g.][]{Carretta2010}. Whereas people have have often
tried to categorize objects as GCs through relations between half-light
radius, density, and surface brightness profile, in fact many objects which are
generally thought of as GCs don't cleanly fit into these cuts
\citep{Peebles1968, Brown1991, Brown1995, Bekki2002}. Consequently,
\citet{Carretta2010} proposed a definition of GCs based on observed chemical
inhomogeneities in their stellar populations. The modern understanding of GCs
then is not simply that of a dense cluster of stars that may have chemical
inhomogeneities and multiple populations; rather, it is one where those
chemical inhomogeneities and multiple populations themselves are the defining
elements of a GC.

All Milky Way globular clusters studied in detail show populations enriched in
He, N, and Na while also being depleted in O and C
\citep{Piotto2015,Bastian2018}. Further, studies of Magellenic Cloud
massive clusters have shown that these light element abundance variations exist
in clusters as young as $\sim 2$ Gyr but not in younger clusters
\citep{Martocchia2019} while there is also evidence of nitrogen variability in
the $\sim 1.5$ Gyr old cluster NGC 1783 \citep{Cadelano2022}.  These light
element abundance patterns also are not strongly correlated with variations in
heavy element abundance, resulting in spectroscopically uniform Fe abundances
between populations \citep[though recent work indicates that there may be
[Fe/H] variations within the first population, e.g.][]{Legnardi2022,
Lardo2022} . Further, high-resolution spectral studies reveal anti-correlations
between N-C abundances, Na-O abundances, and potentially Al-Mg
\citep{Sneden1992, Gratton2012}. Typical stellar fusion reactions can deplete
core oxygen; however, the observed abundances of Na, Al, and Mg cannot be
explained by the CNO cycle \citep{Prantzos2007}. Consequently, globular cluster
populations must be formed by some novel means.

Formation channels for these multiple populations remain a point of debate
among astronomers. Most proposed formation channels consist of some older,
more massive population of stars polluting the pristine cluster media before a
second population forms, now enriched in heavier elements which they themselves could
not have generated \citep[for a detailed review see ][]{Gratton2012}. The four
primary candidates for these polluters are asymptotic giant branch stars
\citep[AGBs,][]{Ventura2001,DErcole2010}, fast rotating massive stars
\citep[FRMSs,][]{Decressin2007}, super massive stars
\citep[SMSs,][]{Denissenkov2014}, and massive interacting binaries
\citep[MIBs,][]{deMink2009, Bastian2018}. 

Hot hydrogen burning (i.e. proton capture), material transport to the surface, and
material ejection into the intra-cluster media are features of each of these
models and consequently they can all be made to {\it qualitatively} agree with
the observed elemental abundances. However, none of the standard models can
currently account for all specific abundances \citep{Gratton2012}. AGB and FRMS
models are the most promising; however, both models have difficulty reproducing
severe O depletion \citep{Ventura2009,Decressin2007}. Moreover, AGB and FRMS
models require significant mass loss ($\sim 90\%$) between cluster formation
and the current epoch --- implying that a significant fraction of halo stars
formed in GCs \citep{Renzini2008,DErcole2008,Bastian2015}.

In addition to the light-element anti-correlations observed, it is also known
that second populations are significantly enhanced in helium
\citep{Piotto2007, Piotto2015, Latour2019}. Depending on the cluster, helium
mass fractions as high as $Y=0.4$ have been inferred \citep[e.g][]{Milone2015}.
However, due to both the relatively high and tight temperature range of partial
ionization for He and the efficiency of gravitational settling in core helium
burning stars, the initial He abundance of globular cluster stars cannot be
observed; consequently, the evidence for enhanced He in GCs originates from
comparison of theoretical stellar isochrones to the observed
color-magnitude-diagrams of globular clusters. Therefore, a careful handling of
chemistry is essential when modeling with the aim of discriminating between
MPs; yet only a very limited number of GCs have been studied with
chemically self-consistent (structure and atmosphere) isochrones
\citep[e.g.][NGC 6752]{Dotter2015}. 

NGC 2808 is the prototype globular cluster to host multiple populations.
Various studies since 2007 have identified that it may host anywhere from two to five 
stellar populations. These populations have been identified both
spectroscopically \citep[i.e.][]{Carretta2004, Carretta2006, Carretta2010,
Gratton2011, Carretta2015, Hong2021} and photometrically
\citep[i.e.][]{Piotto2007, Piotto2015, Milone2015, Milone2017, Pasquato2019}.
Note that recent work \citep{Valle2022} calls into question the statistical
significance of the detections of more than two populations in the spectroscopic
data. Here we present the first stellar structure and evolutionary
models built in a chemically self-consistent manner of NGC 2808.

We model the photometry of the primordial population (hereafter P1) and
the helium enriched population (hereafter P2). \citet{Milone2015} identifies five
populations within NGC 2808, given that the aim of this work is not to identify
sub-populations; rather, to measure the effect that chemical self consistant
stellar structure and evolutionary have on the inferred helium abundance for
the two most extreme cases, we do not consider more than those two
populations. We use archival photometry from the Hubble UV Globular Cluster
Survey (HUGS) \citep{Piotto2015, Milone2017} in the F275W and F814W passbands
to characterize multiple populations in NGC 2808 \citep{Milone2015,
Milone2015b} \citep[This data is available on MAST, ][]{HUGS}. Additionally, we
present a likelihood analysis of the photometric data of NGC 2808 to determine
the number of populations present in the cluster.

\section{Chemical Consistency}\label{sec:const}
There are three primary areas in which the stellar models must be made
chemically consistent: the atmospheric boundary conditions, the opacities, and
interior abundances. The interior abundances are relatively easily handled by
adjusting parameters within our stellar evolutionary code. However, the other
two areas are more difficult to make consistent. Atmospheric boundary
conditions and opacities must both be calculated with a consistent set of
chemical abundances outside of the stellar evolution code.
Nearly all prior efforts at modeling multiple stellar populations in
globular clusters have adjusted the abundances used in the atmospheric interior
models, and in the high temperature opacities, but have not self-consistently
modified the corresponding low-temperature opacities and surface boundary
conditions, as these are found from stellar atmosphere codes, and not the
stellar interior codes which are used to create stellar models and isochrones.
In this work, as in Dotter (2016), the stellar interior models are chemically
self-consistent with the stellar atmosphere models. For evolution, we use the Dartmouth Stellar
Evolution Program (DSEP) \citep{Dotter2008}, a well-tested 1D stellar evolution
code which has a particular focus on modeling low mass stars ($\le 2$
M$_{\odot}$)

\subsection{Atmospheric Boundary Conditions}\label{sec:atm}
Certain assumptions, primarily that the radiation field is at equilibrium and
radiative transport is diffusive \citep{Salaris2005}, made in stellar structure
codes, such as DSEP, are valid when the optical depth of a star is large.
However, in the atmospheres of stars, the number density of particles drops low
enough and the optical depth consequently becomes small enough that these
assumptions break down, and separate, more physically motivated, plasma-
modeling code is required. Generally, structure code will use tabulated
atmospheric boundary conditions generated by these specialized codes, such as ATLAS9
\citep{Kurucz1993}, PHOENIX \citep{Husser2013}, MARCS \citep{Gustafsson2008},
and MPS-ATLAS \citep{Kostogryz2023}. Often, because the boundary conditions are
expensive to compute, they are not updated as interior abundances vary. 

One key element when building chemically self-consistent models of NGC 2808 is the
incorporation of new atmospheric boundary conditions with the same elemental abundances as
the structure code. We use atmospheres generated from the \texttt{MARCS} grid
of model atmospheres \citep{Plez2008}. \texttt{MARCS} provides one-dimensional,
hydrostatic, plane-parallel and spherical LTE atmospheric models
\citep{Gustafsson2008}. Model atmospheres are made to match the
spectroscopically measured elemental abundances of \citet{Milone2015} populations A\&E.
Moreover, for each population, atmospheres with various helium mass fractions
are generated. These range from Y=0.24 to Y=0.36 in steps of 0.03. All
atmospheric models are computed to an optical depth of $\tau = 100$ where their
temperature and pressure serve as boundary conditions for the structure code.
In general, enhancing helium in the atmosphere has only a small impact on the atmospheric
temperature profile, while leading to a drop in the pressure by $\sim 10 - 20 \%$.

\subsection{Opacities}\label{sec:opac}
In addition to the atmospheric boundary conditions, both the high and low
temperature opacities used by DSEP must be made chemically consistent. Here we
use OPLIB high temperature opacity tables \citep{Colgan2016} retrieved using
the TOPS web-interface. Retrival of high termperature opacities is done using
\texttt{pyTOPSScrape}, first introduced in \citet{Boudreaux2023}. Low
temperature opacity tables are retrieved from the Aesopus 2.0 web-interface
\citep{Marigo2009, Marigo2022}. Ideally, these opacities would be the same used
in the atmospheric models. However, the opacities used in the MARCS models are
not publicly available. As such, we use the opacities provided by the TOPS and
Aesopus 2.0 web-interfaces.

\section{Stellar Models}\label{sec:modeling}
We use the Dartmouth Stellar Evolution Program \citep[DSEP, ][]{Dotter2008} to
generate stellar models. DSEP is a one-dimensional stellar evolution code that
includes a mixing length model of convection, gravitational settling, and
diffusion. Using the solar composition presented in \citep{Grevesse2007}
(GAS07), MARCS model atmosphers, OPLIB high temperature opacities, and AESOPUS
2.0 low temperautre opacities we find a solar calibrated mixing length
parameter, $\alpha_{MLT}$, of $\alpha_{MLT} = 1.901$. Abundance
measurments are derived from populations A\&E in \citet{Milone2015} (for P1 and P2 respectivley).

We use DSEP to evolve stellar models ranging in mass from 0.3 to 2.0 solar
masses from the fully convective pre-main sequence to the tip of the red giant
branch. Below 0.7 $M_{\odot}$ we evolve a model every 0.03 $M_{\odot}$ and
above 0.7 $M_{\odot}$ we evolve a model every 0.05 $M_{\odot}$. We evolve
models over a grid of mixing length parameters from $\alpha_{MLT} = 1.0$ to
$\alpha_{MLT} = 2.0$ in steps of 0.1. For each mixing length, a grid of models
and isochrones were calculated, with chemical compositions consistent with
\citet{Milone2015} populations A and E (see Tables \ref{tab:comp} and
\ref{tab:simpleComp}) and a range of helium abundances (Y=0.24, 0.27, 0.30,
0.33, 0.36, and 0.39). In total, 144 sets of isochrones, each with a unique
composition and mixing length were calculated. Each model is evolved in DSEP
with typical numeric tolerences of one part in $10^{7}$. Each model is allowed
a maximum time step of 50 Myr. 

\begin{deluxetable}{c|cc||c|cc}\label{tab:comp}

\tablecaption{Population Composition}

\tablenum{1}

  \tablehead{\colhead{Element} & \colhead{P1 (A)} & \colhead{P2 (E)} & \colhead{Element} & \colhead{P1 (A)} & \colhead{P2 (E)} 
} 

\startdata
Li & -0.08 & --- & In & -1.46 & --- \\
Be & 0.25 & --- & Sn & -0.22 & --- \\
B & 1.57 & --- & Sb & -1.25 & --- \\
C & 6.87 & 5.91 & Te & -0.08 & --- \\
N & 6.42 & 6.69 & I & -0.71 & --- \\
O & 7.87 & 6.91 & Xe & -0.02 & --- \\
F & 3.43 & --- & Cs & -1.18 & --- \\
Ne & 7.12 & 6.7 & Ba & 1.05 & --- \\
Na & 5.11 & 5.7 & La & -0.03 & --- \\
Mg & 6.86 & 6.42 & Ce & 0.45 & --- \\
Al & 5.21 & 6.61 & Pr & -1.54 & --- \\
Si & 6.65 & 6.77 & Nd & 0.29 & --- \\
P & 4.28 & --- & Pm & -99.0 & --- \\
S & 6.31 & 5.89 & Sm & -1.3 & --- \\
Cl & -1.13 & 4.37 & Eu & -0.61 & --- \\
Ar & 5.59 & 5.17 & Gd & -1.19 & --- \\
K & 3.9 & --- & Tb & -1.96 & --- \\
Ca & 5.21 & --- & Dy & -1.16 & --- \\
Sc & 2.02 & --- & Ho & -1.78 & --- \\
Ti & 3.82 & --- & Er & -1.34 & --- \\
V & 2.8 & --- & Tm & -2.16 & --- \\
Cr & 4.51 & --- & Yb & -1.42 & --- \\
Mn & 4.3 & --- & Lu & -2.16 & --- \\
Fe & 6.37 & --- & Hf & -1.41 & --- \\
Co & 3.86 & --- & Ta & -2.38 & --- \\
Ni & 5.09 & --- & W & -1.41 & --- \\
Cu & 3.06 & --- & Re & -2.0 & --- \\
Zn & 2.3 & --- & Os & -0.86 & --- \\
Ga & 0.78 & --- & Ir & -0.88 & --- \\
Ge & 1.39 & --- & Pt & -0.64 & --- \\
As & 0.04 & --- & Au & -1.34 & --- \\
Se & 1.08 & --- & Hg & -1.09 & --- \\
Br & 0.28 & --- & Tl & -1.36 & --- \\
Kr & 0.99 & --- & Pb & -0.51 & --- \\
Rb & 0.26 & --- & Bi & -1.61 & --- \\
Sr & 0.61 & --- & Po & -99.0 & --- \\
Y & 1.08 & --- & At & -99.0 & --- \\
Zr & 1.45 & --- & Rn & -99.0 & --- \\
Nb & -0.8 & --- & Fr & -99.0 & --- \\
Mo & -0.38 & --- & Ra & -99.0 & --- \\
Tc & -99.0 & --- & Ac & -99.0 & --- \\
Ru & -0.51 & --- & Th & -2.2 & --- \\
Rh & -1.35 & --- & Pa & -99.0 & --- \\
Pd & -0.69 & --- & U & -2.8 & --- \\
\enddata
  \tablecomments{Relative Metal composition used where a(H) = 12. Composition
  measurments are taken from \citet{Milone2015} populations A\&E (P1 and P2
  respectively). Where the relative composition is the the same for both P1 and
  P2; it is only listed in the P1 column for the sake of visual clarity.}
\tablerefs{\citet{Milone2015}}
\end{deluxetable}

\begin{deluxetable*}{c|c c c c c c c c c c c}\label{tab:simpleComp}

\tablecaption{Population Abundance Ratios}

\tablenum{2}

  \tablehead{\colhead{Population} & \colhead{[Fe/H]} & \colhead{[$\alpha$/Fe]} & \colhead{[C/Fe]} & \colhead{[N/Fe]} & \colhead{[O/Fe]} & \colhead{[r/Fe]} & \colhead{[s/Fe]} & \colhead{C/O} & \colhead{X} & \colhead{Y} & \colhead{Z} 
} 

\startdata
  A(1) & -1.13 & 0.32 & -0.43 & -0.28 & 0.31 & -1.13 & -1.13 & 0.10 & 0.7285 & 0.2700 & 0.00154 \\
  E(2) & -1.13 & -0.11 & -1.39 & -0.02 & -0.66 & -1.13 & -1.13 & 0.10 & 0.7594 & 0.240 & 0.00063
\enddata
  \tablecomments{Abundance ratios for populations P1 and P2 in NGC 2808.}
\tablerefs{\citet{Milone2015}}
\end{deluxetable*}

For each combination of populations, $Y$, and $\alpha_{MLT}$ we use the
isochrone generation code first presented in \citet{Dotter2016} to generate a
grid of isochrones. The isochrone generation code identified equivalent
evolutionary points (EEPs) over a series of masses and interpolates between
them. The grid of isochrones generated for this work is avalible as a digital
supplement to this paper \dataset[10.5281/zenodo.10631439]{\doi{10.5281/zenodo.10631439}}. Given the complexity of the parameter space when
fitting multiple populations, along with the recent warnings in the literature
regarding overfitting datasets \citep[e.g. ][]{Valle2022}, we want to develop a
more objective way of fitting isochrones to photometry than if we were to mark
median ridge line positions by hand.

\section{fidanka}\label{sec:fidanka}
When fitting isochrones to the clusters with multiple populations we have four
main criteria for any method:

\begin{itemize}
  \item The method must be robust enough to work along the entire main
    sequence, turn off, and much of the subgiant and red giant branch.
	\item Any method should consider photometric uncertainty in the fitting process.
	\item The method should be model independent, weighting any n number of populations equally.
	\item The method should be automated and require minimal intervention from the user.
\end{itemize}

We do not believe that any currently available software is a match for
our use case. Therefore, we have developed our own software suite, \fidanka.
\fidanka is a Python package designed to automate much of the process of
measuring fiducial lines in CMDs, adhering to the four criteria we lay out
above. Primary features of \fidanka may be separated into three
categories: fiducial line measurement, stellar population synthesis, and
isochrone optimization/fitting. Additionally, there are utility functions that
are detailed in the \fidanka documentation.

\subsection{Fiducial Line Measurement}\label{sec:sub:fiducial}
\fidanka takes a iterative approach to measuring fiducial lines, the first step
of which is to make a ``guess'' as to the fiducial line. This initial guess
is calculated by splitting the CMD into magnitude bins, with uniform numbers of
stars per bin (so that bins cover a small magnitude range over densely
populated regions of the CMD, while covering a much larger magnitude range in
sparsely populated regions of the CMD, such as the RGB). A unimodal Gaussian
distribution is then fit to the color distribution of each bin, and the
resulting mean color is used as the initial fiducial line guess. This rough
fiducial line will approximately trace the area of highest density. The initial
guess will be used to verticalize the CMD so that further algorithms can work in
1D magnitude bins without worrying about weighting issues caused by varying
projections of the evolutionary sequence onto the magnitude axis.
Verticalization is performed by taking the difference between the guess fiducial
line and the color of each star in the CMD.

\begin{figure*}
	\centering
	\includegraphics[width=0.9\textwidth]{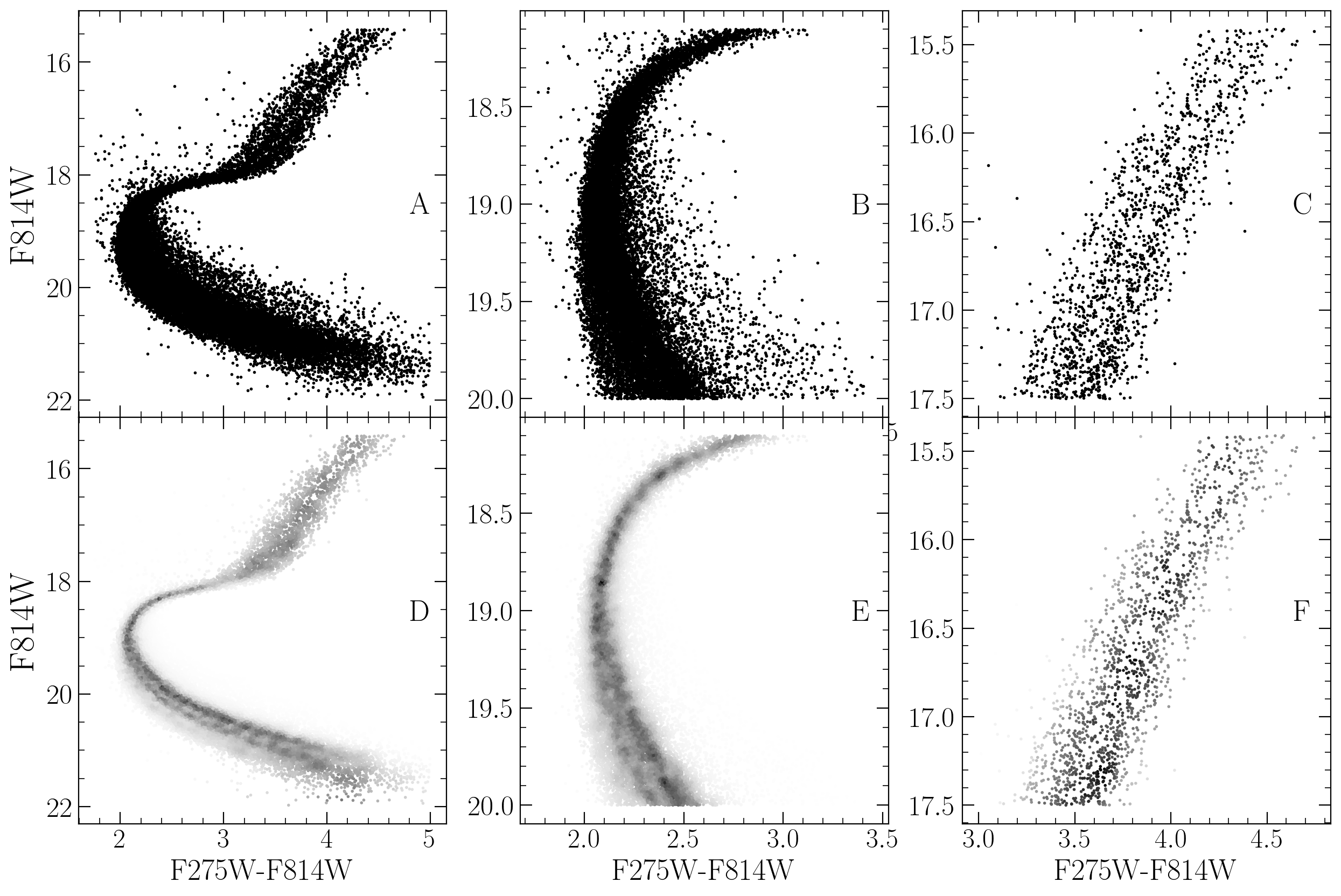}
  \caption{Figures in the top row are the raw CMD, while figures in the bottom
  row are colored by the density map. Density map demo showing density estimate
  over different parts of the evolutionary sequence. The left panel shows the
  density map over the entire evolutionary sequence, while the middle panel
  shows the density map over the main sequence and the right panel shows
  the density map over the RGB. }
	\label{densityMapDemo}
\end{figure*}

If \fidanka were to simply apply the same algorithm to the verticalized CMD,
then the resulting fiducial line would likely be a re-extraction of the initial
fiducial line guess. To avoid this outcome, we take a more robust, number-density based
approach that considers the distribution of stars in both color and magnitude
space simultaneously. As an example, in the case of this work, for
each star in the CMD we first use an \texttt{introselect} partitioning
algorithm to select the 50 nearest stars in F814W vs. F275W-F814W space.
It should be noted that unlike methods using chromosome maps \fidanka
only considers two filters and therefore might lose access to information better
traced by other filters. To account for the case where the star is at an
extreme edge of the CMD, those 50 stars include the star itself (such that we
really select 49 stars + 1). We use
\texttt{qhull}\footnote{https://www.qhull.com}\citep{Barber1996} to calculate
the convex hull of those 50 points. The number density at each star then is
defined as $50/A_{hull}$, where $A_{hull}$ is the area of the convex hull.
Because we use a fixed number of points per star, and a partitioning algorithm
as opposed to a sorting algorithm, this method scales like $\mathcal{O}(n)$,
where n is the number of stars in the CMD. This method also intrinsically
weights the density of each star equally, as the counting statistics per bin
are uniform. We are left with a CMD in which each star has a defined number
density (Figure \ref{densityMapDemo}).

\begin{figure}
	\centering
	\includegraphics[width=0.45\textwidth]{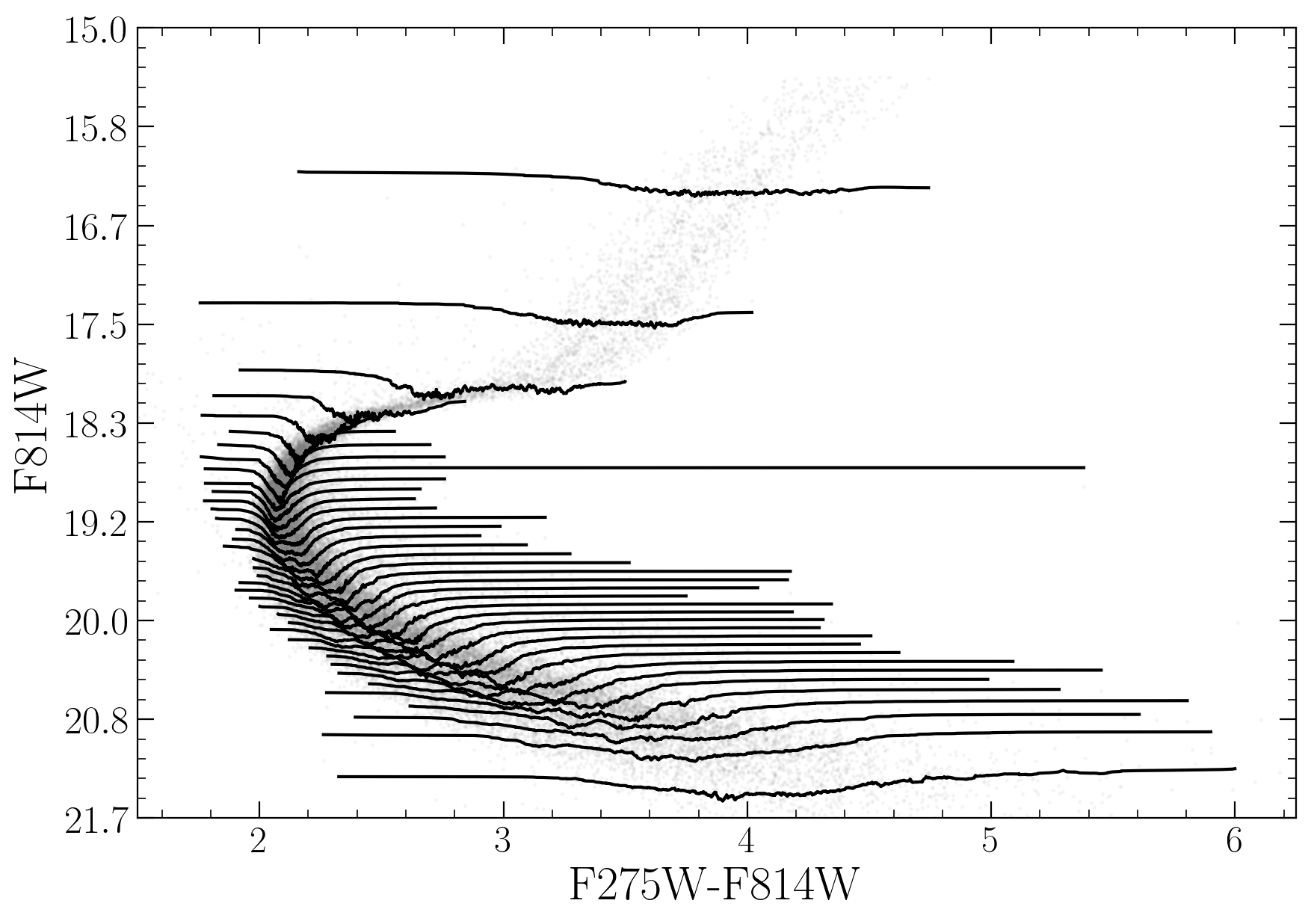}
	\caption{CMD where point brightness is determined by local density. Lines show the
	density-color profile in each magnitude bin. In this figure adaptive
	binning targeted 1000 stars per bin}
	\label{densityBinsDemo}
\end{figure}

\fidanka can now exploit this density map to fit a better fiducial line to the
data, as the density map is far more robust to outliers. There are multiple
algorithms that we implement to fit the fiducial line to the color-density profile
in each magnitude bin (Figure \ref{densityBinsDemo}); these are explained in
more detail in the \fidanka documentation. However, of most relevance here is
the Bayesian Gaussian Mixture Modeling (BGMM) method. BGMM is a clustering
algorithm that, for some fixed number of n-dimensional Gaussian distributions,
$K$, determines the mean, covariance, and mixing probability (somewhat
analogous to amplitude) of each $k^{th}$ distribution, such that the local
lower bound of the likelihood of each star belonging strongly to a single
distribution is maximized. 

\begin{figure*}
	\centering
	\includegraphics[width=0.9\textwidth]{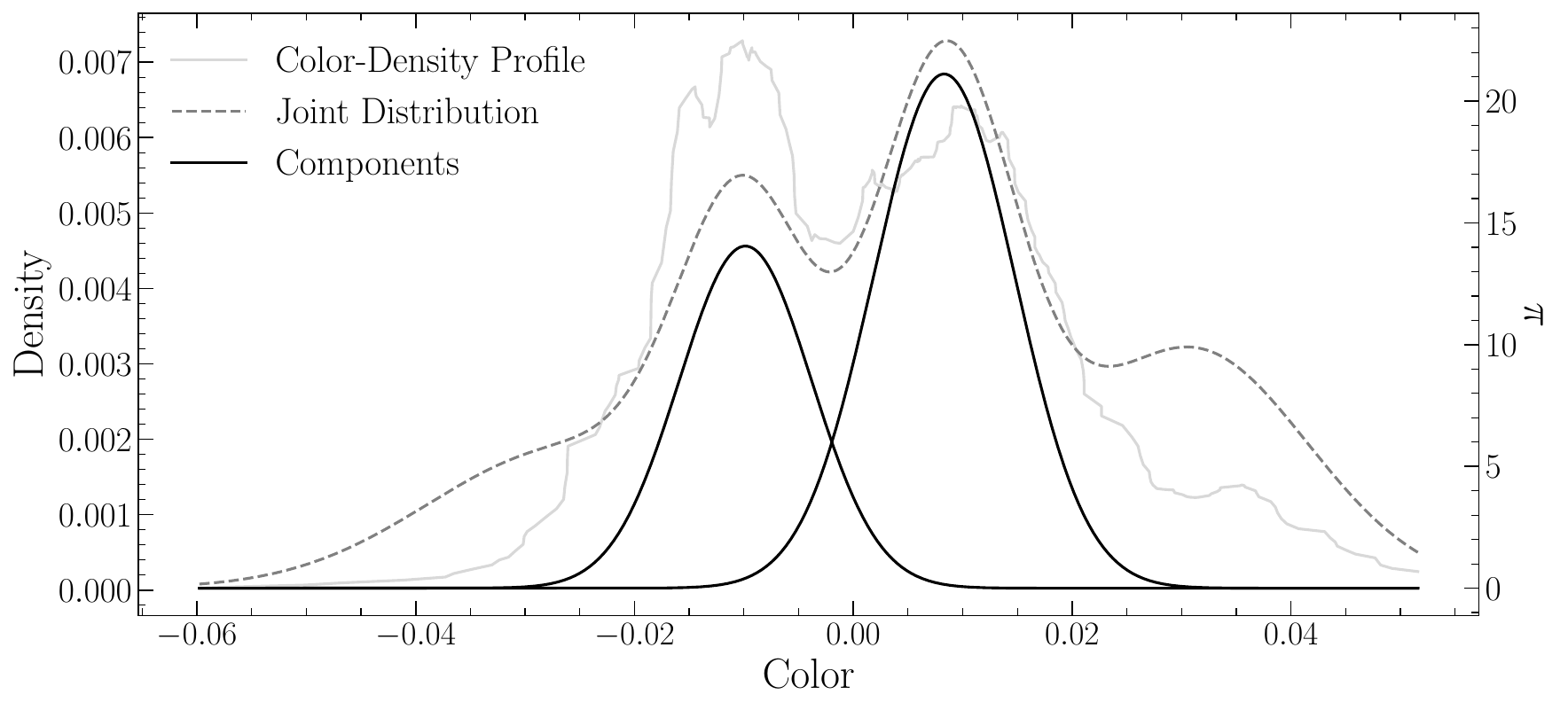}
	\caption{Example of BGMM fit to a magnitude bin. The grey line shows the
	underlying color-density profile, while the black dashed line shows the
	joint distribution of each BGMM component. The solid black lines show the
	two selected components.}
	\label{fig:BGMMDist}
\end{figure*}

Maximization is performed using the Dirichlet process, which is a
non-parametric Bayesian method of determining the number of Gaussian
distributions, $K$, that best fit the data \citep{Ferguson1973, scikit-learn}.
Use of the Dirichlet process allows for dynamic variation in the number of
inferred populations from magnitude bin to magnitude bin. Specifically,
populations are clearly visually separated from the lower main sequence through
the turn off; however, at the turn off and throughout much of the subgiant
branch, the two visible populations overlap due to their similar ages
\citep[i.e.][]{Jordan2002}. The Dirichlet process allows for the BGMM method to
infer a single population in these regions, while inferring two populations in
regions where they are clearly separated. More generally, the use of the
Dirichlet process removes the need for a prior on the exact number of
populations to fit. Rather, the user specifies a upper bound on the number of
populations within the cluster. An example bin (F814W = 20.6) is shown in
Figure \ref{fig:BGMMDist}.

\begin{figure}
	\centering
	\includegraphics[width=0.45\textwidth]{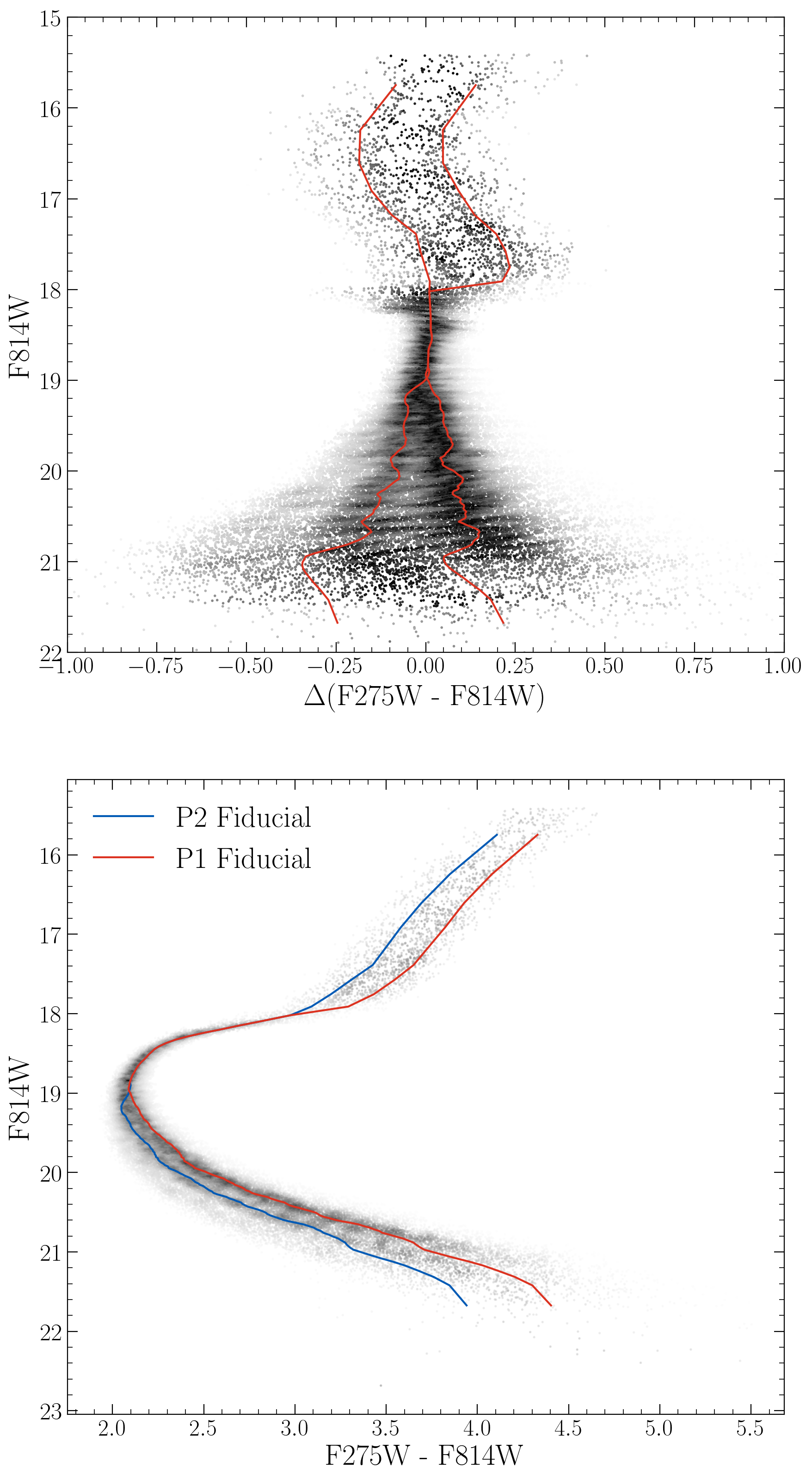}
  \caption{Verticalized CMD (where the color of each data point is subtracted
  from the color of the fiducial line at that magnitude) where point brightness
  is determined by density (top). CMD where point brightness is determined by
  density, calculated fiducial lines are shown (bottom). The data used is from
  the Hubble Space Telescope UV Legacy Survey of Galactic Globular Clusters.}
	\label{fig:vertFit}
\end{figure}

\fidanka's BGMM method first breaks down the verticalized CMD into magnitude
bins with uniform numbers of stars per bin (here we adopt 250). Any stars left
over are placed into the final bin. For each bin a BGMM model with a maximum of
five populations is fit to the color density profile. The number of populations is
then inferred from the weighting parameter (the mixing probability) of each
population. If the weighting parameter of any $k^{th}$ components is less than
{\color{blue}0.05}, then that component is considered to be spurious and
removed. Additionally, if the number of populations in the bin above and the
bin below are the same, then the number of populations in the current bin is
forced to be the same as the number of populations in the bin above. Finally,
the initial guess fiducial line is added back to the BGMM inferred line. Figure
\ref{fig:vertFit} shows the resulting fiducial line(s) in each magnitude bin
for both a verticalized CMD and a non-verticalized CMD. In contrast to other
work in the literature where evidence for up to five distinct populations has been
found, we only find evidence for two stellar populations.

This method of fiducial line extraction effectively discriminated between
multiple populations along the main sequence and RGB of a cluster, while
simultaneously allowing for the presence of a single population along the MSTO
and subgiant branch. 

We can adapt this density map-based BGMM method to consider photometric
uncertainties by adopting a simple Monte Carlo approach. Instead of measuring
the fiducial line(s) a single time, \fidanka can measure the fiducial line(s)
many times, resampling the data with replacements each time. For each resampling,
\fidanka adds a random offset to each filter based on the photometric
uncertainties of each star. From these $n$ measurements the mean fiducial line
for each sequence can be identified along with upper and lower-bound confidence
intervals in each magnitude bin.

\subsection{Stellar Population Synthesis}
While not extensively used in this paper \fidanka can, in addition to measuring fiducial
lines, perform stellar population synthesis. \fidanka's population synthesis
module can generate synthetic stellar populations from a set of MIST-formatted
isochrones. This is of primary importance for binary population modeling. The
module is also used to generate synthetic CMDs for the purpose of testing the
fiducial line extraction algorithms against priors.

\fidanka uses MIST-formatted isochrones \citep{Dotter2016} as input along
with distance modulus, B-V color excess, binary mass fraction, and bolometric
corrections. An arbitrarily large number of isochrones may be used to define an
arbitrary number of populations. Synthetic stars are samples from each
isochrone based on a definable probability; For example, it is believed that
$\sim90\%$ of stars in globular clusters are younger population
\citep[e.g.][]{Suntzeff1996, Carretta2013}. Based on the metallicity, $\mu$, and E(B-V) of each
isochrone, bolometric corrections are taken from bolometric correction tables.
Where bolometric correction tables do not include exact metallicities or
extinctions a linear interpolation is performed between the two bounding
values. 

\subsection{Isochrone Optimization}
The optimization routines in \fidanka will find the best fit distance modulus,
B-V color excess, and binary number fraction for a given set of isochrones. If
a single isochrone is provided then the optimization is done by minimizing the
$\chi^2$ of the perpendicular distances between an isochrone and a fiducial
line. If multiple isochrones are provided then those isochrones are first used
to run a stellar population synthesis and generate a synthetic CMD. The
optimization is then done by minimizing the $\chi^2$ of both the perpendicular
distances between and widths of the observed fiducial line and the fiducial
line of the synthetic CMD.

\subsection{Fidanka Testing}
\begin{figure}
  \centering
  \includegraphics[width=0.45\textwidth]{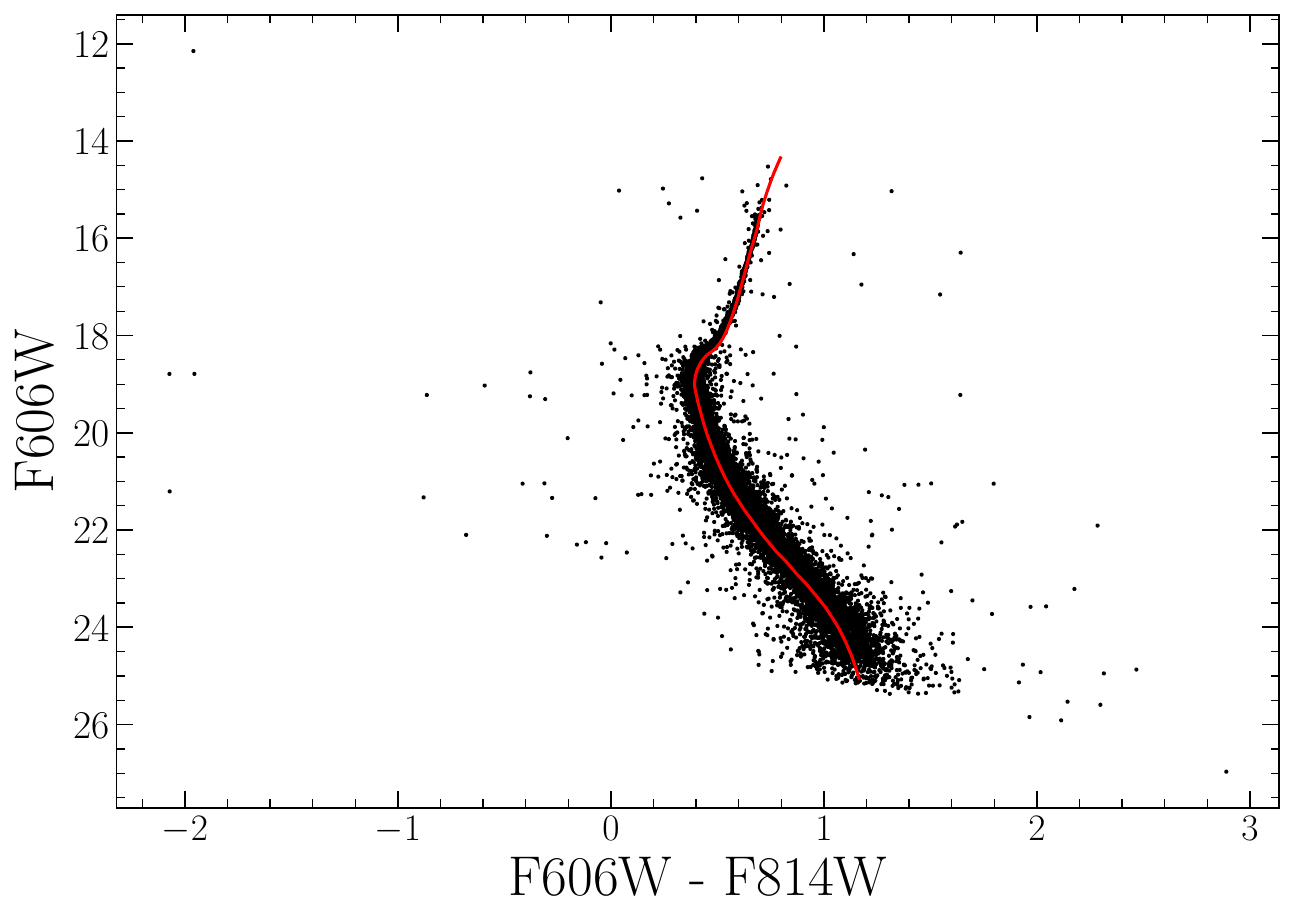}
  \caption{Synthetic population generated by \fidanka at 10000pc with E(B-V) =
  0, and an age of 12 Gyr along with the best fitting isochrone. The best fit
  paremeters are derived to be $\mu=15.13$, E(B-V)=0.001, and an age of 12.33
  Gyr.}
  \label{fig:ValidationBestFit}
\end{figure}

In order to validate \fidanka we have run a series of injection recovery tests
using \fidanka's population synthesis routines to build various synthetic
populations and \fidanka's fiducial measurement routines to recover these
populations. Each population was generated using the initial mass function
given in \citep{Milone2012} for the redmost population ($\alpha=-1.2$).
Further, every population was given a binary population fraction of 10\%,
distance uniformly sampled between 5000pc and 15000pc, and a B-V color excess
uniformly sampled between 0 and 0.1. \fidanka makes use of ACS
artificial star tests \citep{Anderson2008} to model the noise and
completness of a synthetic population in passbands covered by those tests. Full
details on how \fidanka uses artificial star tests may be found on its
documentation page\footnote{https://tboudreaux.github.io/fidanka/} Finally,
each synthetic population was generated using a fixed age uniformly sampled
between 7 Gyr and 14 Gyr. An example synthetic population, along with its
associated best fit isochrone, are
shown in Figure \ref{fig:ValidationBestFit}.

\begin{figure}
  \centering
  \includegraphics[width=0.45\textwidth]{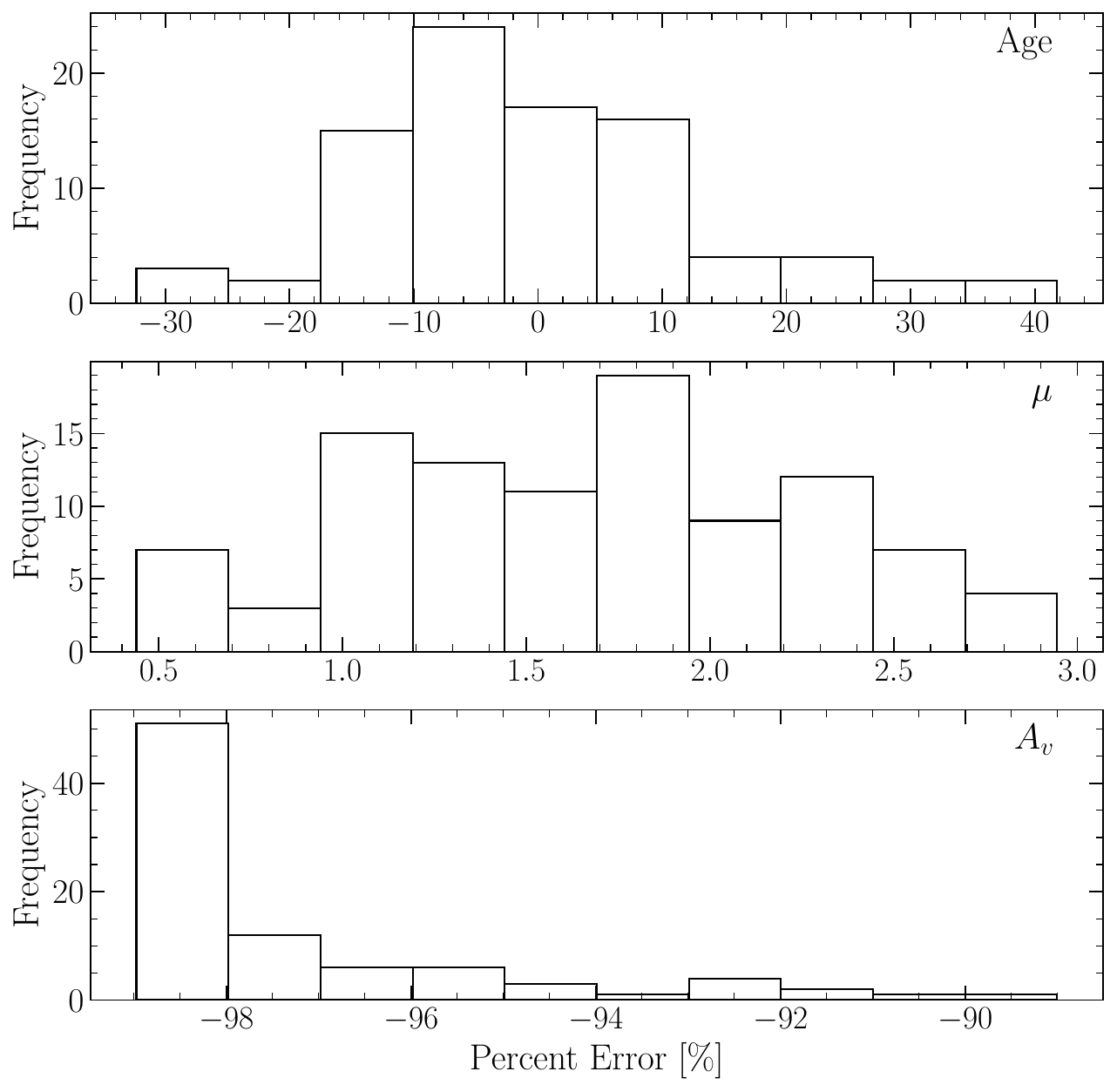}
  \caption{Percent error distribution for each of the three deriver parameters.
  Note that these values will be sensitive to the magnitude uncertainties of
  the photometry. Here we made use of the ACS artificial star tests to estimate
  the uncertanties.}
  \label{fig:validationDist}
\end{figure}

For each trial we use \fidanka to measure the fiducial line and then optimize
that fiducial line against the originating isochrone to esimate distance
modulus, age, and color B-V excess. Figure \ref{fig:validationDist} is built
from 1000 Monte-Carlo trials and shows the mean and width of the percent
error distributions for $\mu$, $A_{v}$, and age. In general \fidanka is able to
recover distance modulii effectively with age and E(B-V) recovery falling in
line with other literature that does not cosider the CMD outside of the main
sequence, main sequence turn off, subgiant, and red giant branches.
 Specifically, it should be noted that \fidanka is not set up to model the
horizontal branch.

\section{Isochrone Fitting}\label{sec:isoFit}
We fit pairs of isochrones to the HUGS data for NGC 2808 using \fidanka, as
described in \S \ref{sec:fidanka}. As was mentioned in \S
\ref{sec:sub:fiducial}, the method used by \fidanka only consideres two filters --- in
the case of this work F275W and F814W --- and therefore might be unable to
distinguish between populations separated only in the higher-dimensional space
of a chromosome map. For further discussion of why we adopt this method, despite it limits, we refer the reader to \S \ref{sec:sub:numpops}. Two isochrones, one for P1 and one for P2 are fit
simultaneously. These isochrones are constrained to have distance modulus,
$\mu$, and color excess, E(B-V) which agree to within 0.5\% and an ages which
agree to within 1\%. Moreover, we constrain the mixing length, $\alpha_{ML}$,
for any two isochrones in a set to be within 0.5 of one and other. For each 
isochrone set we optimize $\mu_{P1}$, $\mu_{P2}$, $E(B-V)_{P1}$, $E(B-V)_{P2}$, Age$_{P1}$, and $Age_{P2}$ in order to reduce the
$\chi^{2}$ distance ($\chi^{2} = \sum\sqrt{\Delta \text{color}^{2} + \Delta
\text{mag} ^{2}}$) between the fiducial lines and the isochrones. Because we
fit fiducial lines directly, we do not need to consider the binary population
fraction, $f_{bin}$, as a free parameter.

\begin{figure*}
  \centering
  \includegraphics[width=0.9\textwidth]{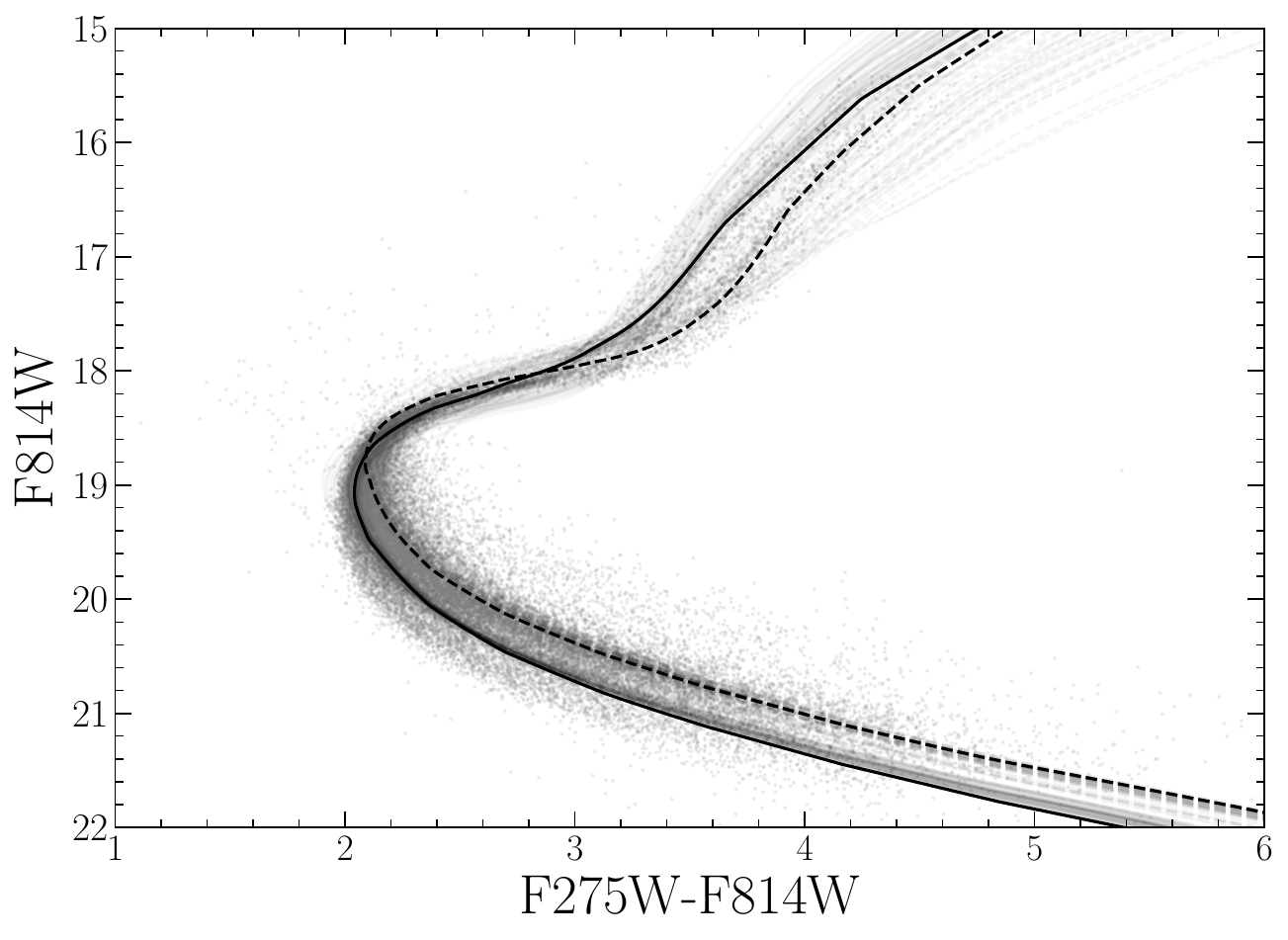}
  \caption{Best fit isochrone results for NGC 2808. The best fit P1 and P2
  models are shown as black lines. The following 50 best fit models are
  presented as grey lines. The dashed black line is fit to P1, while
  the solid black line is fit to P2.}
  \label{fig:BestFitResults}
\end{figure*}

The best fit isochrones are shown in Figure \ref{fig:BestFitResults} and
optimized parameters for these are presented in Table \ref{tab:BestFitResults}.
The initial guess for the age of these populations was locked to 12 Gyr
and the initial extinction was locked to 0.5 mag. The initial guess for the
distance modulus was determined at run time using a dynamic time warping
algorithm to best align the morphologies of the fiducial line with the target
isochrone. This algorithm is explained in more detail in the \fidanka
documentation under the function called \texttt{guess\_mu}. We find helium mass
fractions that are consistent with those identified in past literature
\citep[e.g.][]{Milone2015}. Note that our helium mass fraction grid has a
spacing of 0.03 between grid points and we are therefore unable to resolve
between certain proposed helium mass fractions for the younger sequence (for
example between 0.37 and 0.39). We also note that the best fit mixing
length parameters which we derive for P1 and P2 do not agree within
their uncertainties. This is not surprising, as the much higher mean molecular mass
of P2 --- when compared to P1, due to population P2's larger
helium mass fraction --- will result in a steeper adiabatic temperature
gradient. 

\begin{table*}
  \centering
  \begin{tabular}{c | c c c c c c}
    \hline
    Population & Age & Distance Modulus & Extinction & Y & $\alpha_{ML}$ & $\chi^{2}_{\nu}$\\
    & [Gyr] & & [mag] & & &\\
    \hline
    \hline
    P1 & 12.996$^{+0.87}_{-0.64}$ & 15.021 & 0.54 & 0.24 & 2.050 & 0.021\\
    P2 & 13.061$^{+0.86}_{-0.69}$ & 15.007 & 0.537 & 0.39 & 1.600 & 0.033 \\
    \hline
  \end{tabular}
  \caption{Best fit parameters derived from fitting isochrones to the fiducual
  lines derived from the NCG 2808 photometry. The one sigma uncertainty
  reported on population age were determined from the 16th and 84th percentiles
  of the distribution of best fit isochrones ages.}
  \label{tab:BestFitResults}
\end{table*}

Past literature \citep[e.g. ][]{Milone2015, Milone2018} has found helium mass
fraction variation from the redmost to bluemost populations of $\sim 0.12$.
Here we find a helium mass fraction variation of 0.15 that, given the spacing
of the helium grid we use {\em is consistent with these past results}.
The helium mass fractions we derive for P1 and P2 are consistent with
those of populations A and E in \cite{Milone2015}; however, populations B+C and
D in that study are more prominently separated in the F275W-F814W colorband.
The inferred helium mass fractions for P1 and P2 are not consistent  
with those reported for populations B+C and D.

\subsection{The Number of Populations in NGC 2808}\label{sec:sub:numpops} 
\begin{figure}
  \centering
  \includegraphics[width=0.45\textwidth]{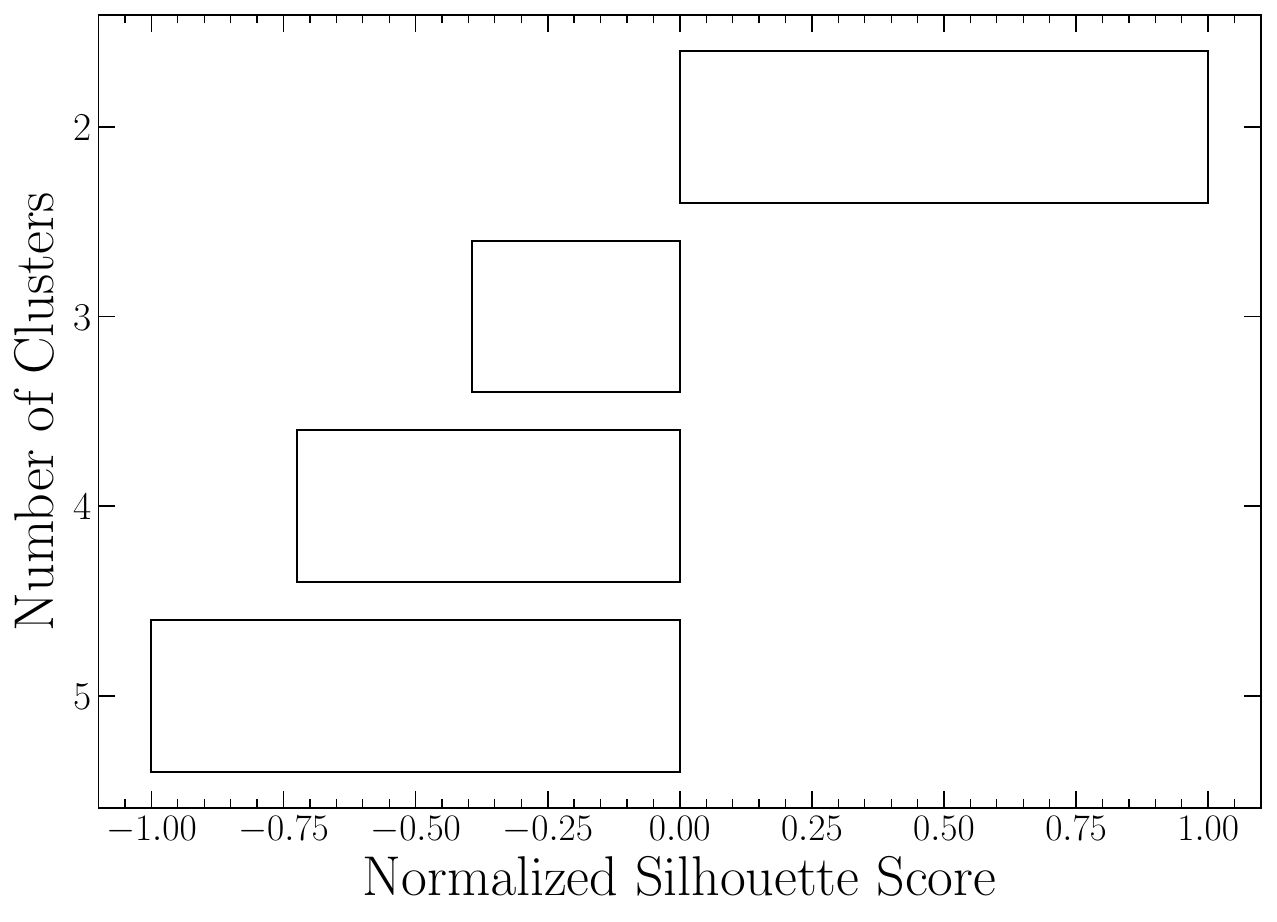}
  \includegraphics[width=0.45\textwidth]{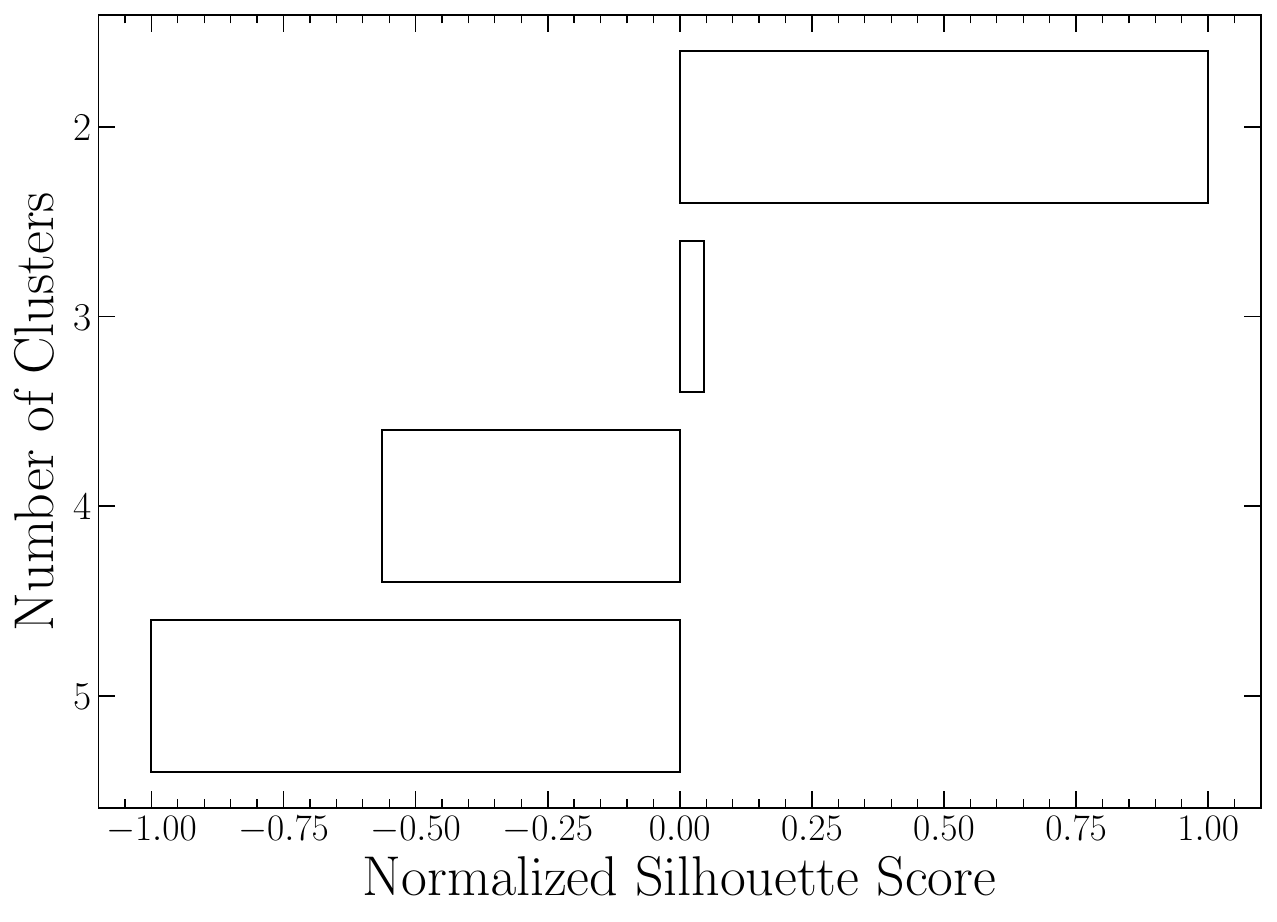}
  \caption{Silhouette analysis for NGC 2808 F275W-F814W (top) and F336W-F438W
  (bottom) photometry. The Silhouette scores are an average of score for each
  magnitude bin. Scores have been normalized to indicate the most
  well-distinguished (+1) to least well-distinguished (-1) hypothesizes.}
  \label{fig:clusterAn}
\end{figure}

In
order to estimate the number of populations that ideally fit the NGC 2808
F275W-F814W photometry without overfitting the data we make use of silhouette
analysis \citep[][and in a similar manner to how \citet{Valle2022} perform
their analysis of spectroscopic data]{ROUSSEEUW198753, Shahapure2020}. We find
the average silhouette score for the hypothesizes that there are two, three,
four, or five population in each magnitude bin. We preform this analysis over
all magnitude using routines built into the standard Python module
\texttt{sklearn}. Figure \ref{fig:clusterAn} (top) shows the silhouette
analysis results and that two populations fit the photometry most ideally. This
result is in line with what our BGMM model predicts for the majority of the
CMD.

\begin{figure}
  \centering
  \includegraphics[width=0.45\textwidth]{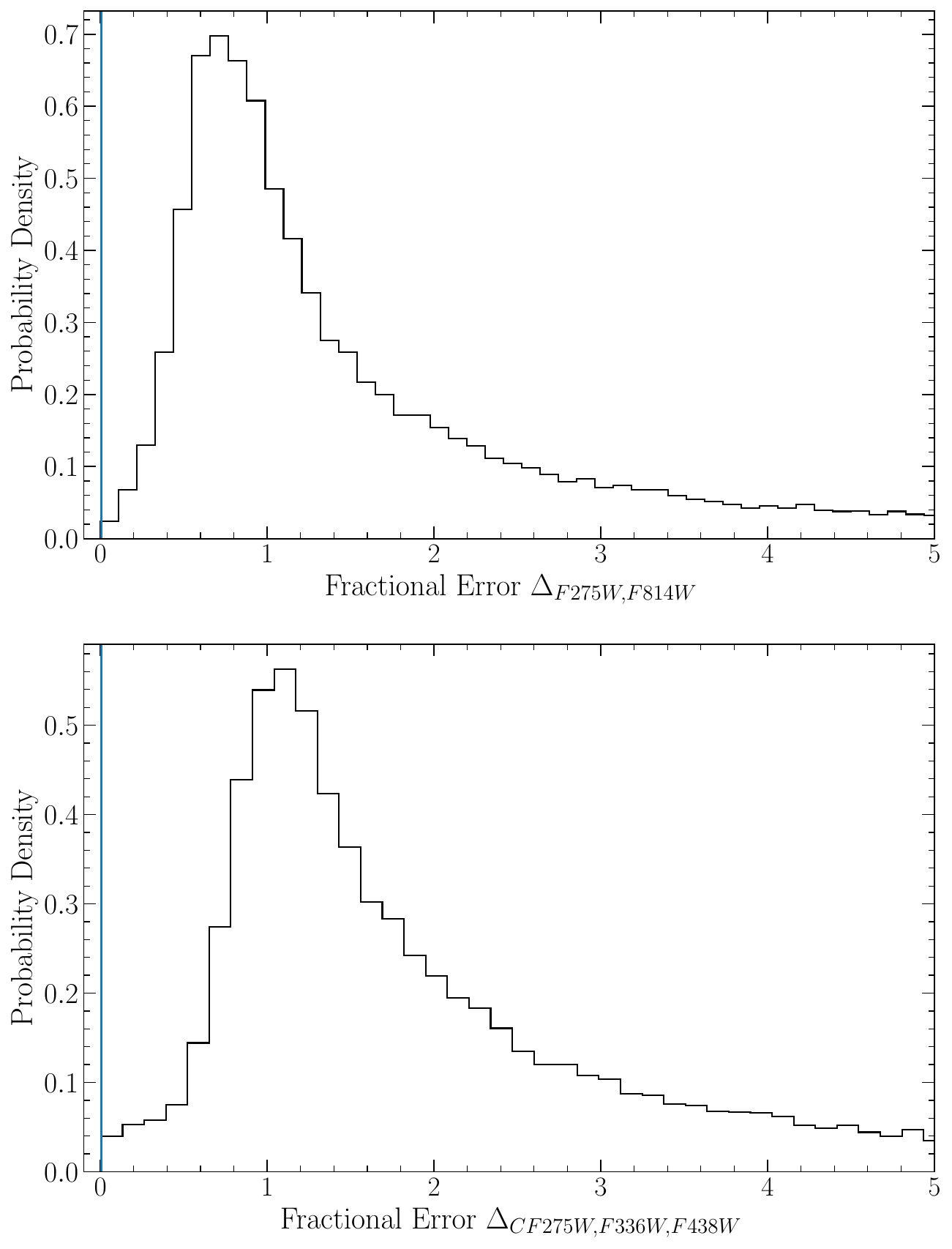}
  \caption{Fractional error distribution of $\Delta_{F275W,F814W}$
  (top) and $\Delta_{CF275W,F336W,F438W}$. The vertical line near 0 in each
  figure indicates the mean fractional error of the magnitude measurements used
  to find the chromosome parameters.}
  \label{fig:chromosomeError}
\end{figure}

While we make use of a purely CMD-based approach in this work, other
literature has made use of chromosome maps. These consist of implicitly
verticalized pseudo colors. In the chromosome map for NGC 2808 there may be
evidence for more than two populations; further, the chromosome maps used
include information from additional filters (F336W and F438W) which we do not
use in our CMD approach. We preform the same analysis on the F336W-F438W CMD
using \fidanka as we do on the F275W-F814W CMD. While the clustering algorithm
does find a more strongly distinguished potential third population  using these
filters (Figure \ref{fig:clusterAn} bottom), the two population hypothesis remains strongly preferred. Moreover, the
process of transforming magnitude measurements into chromosome space results in
dramatically increased uncertainties for each star. We find a mean fractional
uncertainty for chromosome parameters --- $\Delta_{F275W,F814W}$ and
$\Delta_{CF275W,F336W,F438W}$ --- of $\approx1$ (Figure
\ref{fig:chromosomeError}) when starting with magnitude measurements having a
mean best-case (i.e. where the uncertainty is assumed to only be due to Poisson statistics)
fractional uncertainty of $\approx 0.0005$. Fractional uncertainties for
chromosome parameters were calculated via standard propagation of uncertainty.
Because of how \fidanka operates, i.e. resampling a probability distribution for
each star in order to identify clusters, we are unable to make statistically
meaningful statements from the chromosome map

\section{Conclusion}\label{sec:conclusion}
Here we have performed the first chemically self-consistent modeling of the
Milky Way Globular Cluster NGC 2808. We find that, updated atmospheric boundary
conditions and opacity tables do not have a significant effect on the inferred
helium abundances of multiple populations. Specifically, we find that
P1 has a helium mass fraction of 0.24, while P2 has a helium
mass fraction of 0.39. Additionally, we find that the ages of these two populations 
agree within uncertainties. We only find evidence for two distinct stellar
populations, which is in agreement with recent work studying the number
of populations in NGC 2808 spectroscopic data.

We introduce a new software suite for globular cluster science,
\fidanka, which has been released under a permissive open source license.
\fidanka aims to provide a statistically robust set of tools for estimating the
parameters of multiple populations within globular clusters.

\acknowledgments{
  This work has made use of the NASA astrophysical data system (ADS). We would
  like to thank Elisabeth Newton and Aaron Dotter for their support and for
  useful discussion related to the topic of this paper. We would like to thank
  the reviewers of this work for their careful reading and critique which has
  significantly improved this article. We would like to thank
  Kara Fagerstrom, Aylin Garcia Soto, and Keighley Rockcliffe for their useful
  discussion related to in this work. We acknowledge the support of a NASA
  grant (No. 80NSSC18K0634). Emily M. Boudreaux additionally received financial support from
  the European Research Council (ERC) under the Horizon Europe programme
  (Synergy Grant agreement No. 101071505: 4D-STAR). Work for this review was
  partially funded by the European Union. Views and opinions expressed are
  however those of the author(s) only and do not necessarily reflect those of
  the European Union or the European Research Council. Neither the European
  Union nor the granting authority can be held responsible for them.
}
\acknowledgments


\bibliography{ms}{}
\bibliographystyle{aasjournal}

\end{document}